\documentstyle[12pt]{article}
\newcommand{\be}{\begin{equation}}
\newcommand{\ee}{\end{equation}}
\newcommand{\bea}{\begin{eqnarray}}
\newcommand{\eea}{\end{eqnarray}}

\newcounter{saveeqn}
\newcommand{\aeqn}
   {\stepcounter{equation}
    \setcounter{saveeqn}{\value{equation}}
    \setcounter{equation}{0}
    \renewcommand{\theequation}{\mbox{\arabic{saveeqn}\alph{equation}}}}
\newcommand{\reqn}
   {\setcounter{equation}{\value{saveeqn}}
    \renewcommand{\theequation}{\arabic{equation}}}
\title{
{\vspace{-3cm} \normalsize \hfill
                        \parbox{35mm}{MS-TPI-94-01 \\
                                      hep-lat/9402017 }  }\\[25mm]
Field Theoretic Calculation of the Universal\\
Amplitude Ratio of Correlation Lengths\\
in $3D$-Ising Systems
}

\author{Gernot M\"unster and Jochen Heitger\\
        Institut f\"ur Theoretische Physik I,
        Universit\"at M\"unster\\
        Wilhelm-Klemm-Str.~9, D-48149 M\"unster, Germany}
\date{February 25, 1994}
\begin{document}
\maketitle

\begin{abstract}
In three-dimensional systems of the Ising universality class the ratio
of correlation length amplitudes for the high- and low-temperature
phases is a universal quantity.
Its field theoretic determination apart from the $\epsilon$-expansion
represents a gap in the existing literature.
In this article we present a method, which allows to calculate this
ratio by renormalized perturbation theory in the phases with unbroken
and broken symmetry of a one-component $\phi^4$-theory in fixed
dimensions $D=3$.
The results can be expressed as power series in the renormalized
coupling constant of either of the two phases, and with the knowledge of
their fixed point values numerical estimates are obtainable.
These are given for the case of a two-loop calculation.
\end{abstract}
%
\section{Introduction}
The application of field theoretic methods to critical phenomena
and second order phase transitions is one of the main tools
for their quantitative theoretical investigation.
Based on general theoretical arguments it is believed that
physical systems, which undergo a second order phase transition,
fall into distinct universality classes characterized by the
dimensionality $D$ of space, the number $n$ of components of the
order parameter and the underlying symmetry group.
This manifests itself in the fact that within these universality
classes many interesting quantities such as critical
exponents and amplitude ratios have the same values \cite{A84,BD92}.

As a consequence of the long range nature of the effective interaction
(for example between the spin variables of a ferromagnetic system
showing a collective behaviour in the vicinity of the critical
temperature $T_c$) the microscopic structure of the given system
can be neglected by introducing a real valued order parameter field
$\phi(x)=(\phi_1(x),\ldots,\phi_n(x))$, $x\in{\bf R}^D$,
which is assumed to describe the phase transition.
In this way one is led to a field theory with $\phi^4$-self-interaction
with spontaneous breaking of its $O(n)$-symmetry, which is one of the
standard topics in quantum field theory.

In the past many perturbative calculations have been made to estimate
the values for the above mentioned critical quantities.
To this end the $\epsilon$-expansion, initiated by M.E.~Fisher and
K.G.~Wilson \cite{WF72} and subsequently elaborated by E.~Br\'{e}zin,
J.C.~Le Guillou, J.~Zinn-Justin \cite{BLZ76} and others for the massless
$\phi^4$-theory in $D=4-\epsilon$ dimensions, has been applied
and yielded very good results.

A quantity whose numerical value is important in comparing various
experimentally determined universal quantities with theoretical
predictions (see e.g.\ \cite{M90}) is the universal ratio of correlation
length amplitudes $f_+ / f_-$.
It is defined by the behaviour of the correlation length $\xi$ as a
function of the temperature $T$ near $T=T_c$ through
\be
\xi\sim\left\{\begin{array}{l@{,\quad}l}
f_+t^{-\nu}&t>0\\f_-(-t)^{-\nu}&t<0\end{array}\right.;
\quad t:=\frac{T-T_c}{T_c}\,,
\ee
where the symbol $\sim$ denotes the critical behaviour as defined in the
usual way \cite{A84,BD92}.
The $\epsilon$-expansion of $f_+ / f_-$ is known up to second order for
the second-moment correlation length and up to first order for the
`true' correlation length \cite{BLZ74,BLZ76}, which limits the accuracy
of numerical estimates.

In this article we follow an idea owing to G.~Parisi \cite{Pa80},
who suggested the use of renormalized massive perturbation theory
in fixed dimensions $D=3$.
For the symmetric phase there exist extensive calculations by
G.A.~Baker, B.G.~Nickel, D.I.~Meiron et al.\ \cite{BNM76,BNM77,BNM78},
who give the renormalization group functions and the critical exponents
respectively up to six- and seven-loop order.
Later C.~Bagnuls, C.~Bervillier et al.\ \cite{BB85,BB87} have
developed a renormalization scheme, which among others makes it
possible to determine the amplitude ratios of the susceptibility
and the specific heat from these series.
In general the quality of results from this method appears to be better
than from the $\epsilon$-expansion.

Up to now an extension of this method to the universal amplitude
ratio $f_+/f_-$ is not found in the literature since this requires
explicit calculations in the phase of broken symmetry, too.

In order to fill this gap we consider Euclidean $\phi^4$-theory
with $D=3$ and $n=1$, which is a super-renormalizable field theory,
that is believed to lie in the same universality class as the
three-dimensional Ising model.
Its Lagrangian density in the symmetric phase ($t>0$) is given by
\bea
{\cal L}(\phi_{0^+})&=&\frac{1}{2}\,(\partial\phi_{0^+}(x))^2
                       +{\cal V}(\phi_{0^+})\nonumber\\
{\cal V}(\phi_{0^+})&=&\frac{1}{2}\,m_{0^+}^2\phi^2_{0^+}(x)
     +\frac{1}{4!}\,g_0\phi^4_{0^+}(x)
\label{eq20}
\eea
with $g_0>0$ and $m^2_{0^+}>0$.
The quantities in different phases are distinguished by indices $+$ and
$-$, if necessary.
In the broken phase ($t<0$) the $Z_2$-symmetry
$\phi\rightarrow -\phi$ is lost for the Lagrangian
\bea
{\cal L}(\phi)&=&\frac{1}{2}\,(\partial\phi(x))^2
                 +{\cal V}(\phi)\nonumber\\
{\cal V}(\phi)&=&-\frac{1}{4}\,m^2_{0^-}\phi^2(x)
                 +\frac{1}{4!}\,g_0\phi^4(x)
                 +\frac{3}{8}\,\frac{m^4_{0^-}}{g_0}
                 =\frac{1}{4!}\,g_0\Big(\phi^2(x)-v^2_0\Big)^2
\label{eq30}
\eea
with $m^2_{0^-}>0$ and the classical potential minima at
$\phi=\pm v_0:=\pm\sqrt{3m^2_{0^-}/g_0}$.
After an expansion of the potential around the positive minimum,
which is equivalent to a shift in the field variable via
\be
\phi_{0^-}(x):=\phi(x)-v_0,
\ee
we arrive at
\bea
{\cal L}(\phi_{0^-})&=&\frac{1}{2}\,(\partial\phi_{0^-}(x))^2
                       +{\cal V}(\phi_{0^-})\nonumber\\
{\cal V}(\phi_{0^-})&=&\frac{1}{2}\,m^2_{0^-} \phi^2_{0^-}(x)
     +\frac{1}{3!}\sqrt{3g_0}\,m_{0^-}\phi^3_{0^-}(x)
     +\frac{1}{4!}\,g_0\phi^4_{0^-}(x)\,.
\label{eq50}
\eea
These Lagrangian densities are the starting points of our perturbative
calculations in both phases.
It should be noted that the additional $\phi^3$-self-interaction
of the field $\phi_{0^-}$ in the broken phase gives rise to
tadpole-diagrams as well as to a non-vanishing
one-point function, which is the vacuum expectation value of
the field $\phi_{0^-}$.

The Feynman rules in momentum space follow from the Lagrangians
(\ref{eq20}) and (\ref{eq50}) as usual.
In particular each four-vertex is associated with a factor $-g_0$,
each three-vertex gets a factor $-\sqrt{3g_0}\,m_{0^-}$ and
the Feynman propagator is
\be
\tilde{\triangle}(k):=\frac{1}{k^2+m^2_{0^{\pm}}}\,.
\ee
The divergences of the theory, appearing in the form of the only
two primitively divergent graphs of the two-point function,
are isolated by dimensional regularization.
%
\section{Renormalized perturbation theory}
Before introducing the renormalization schemes for the phases with
unbroken and broken symmetry let us briefly outline our further
strategy.

To begin with we define the correlation length for general
$D$ as the second moment
\be
\xi^2:=\frac{1}{2D}\,\frac{\int d^Dx\,x^2 G^{(2,0)}_c(x)}
                          {\int d^Dx\,G^{(2,0)}_c(x)}
      =-\frac{\frac{\partial}{\partial p^2}\,G^{(2,0)}_c(p)}
             {G^{(2,0)}_c(p)}\,\Bigg|_{p^2=0}
\ee
of the connected two-point correlation function
\be
G^{(2,0)}_c(x):=\langle\phi_0(x)\phi_0(y)\rangle
               -\langle\phi_0(x)\rangle\langle\phi_0(y)\rangle\,.
\ee
Because of its relation
\be
-\Gamma_0^{(2,0)}(p)=\Big(G^{(2,0)}_c(p)\Big)^{-1}
\ee
to the two-point vertex function $\Gamma_0^{(2,0)}$ in momentum space
one verifies the following identity between the correlation length
and the renormalized mass $m_R$:
\be
m^2_R:=\frac{\Gamma_0^{(2,0)}(p)}{\frac{\partial}{\partial p^2}\,
             \Gamma_0^{(2,0)}(p)}\,\Bigg|_{p^2=0}
      =\frac{1}{\xi^2}\,.
\label{eq100}
\ee
As $\Gamma_0^{(2,0)}(p)$ is given by
\be
-\Gamma_0^{(2,0)}(p)=\tilde{\triangle}^{-1}(p)-\Sigma(p)\,,
\ee
where $\Sigma(p)$ is the sum of all one-particle irreducible
two-point graphs with amputated external legs, equation (\ref{eq100})
shows how $\xi$ can be determined perturbatively.
Alternatively one defines the `true' correlation length by the
exponential decay of the two-point function. Then $\xi$ equals
the inverse physical mass (see e.g.\ \cite{MM94}).

{}From renormalization group theory the critical behaviour of the model
is known to be controlled by the dimensionless renormalized coupling
constant $u_R$ and its non-trivial, infrared-stable fixed point $u_R^*$.
The calculation of renormalized quantities sketched in this
section enables us to find analytic functions in $u_R$, in terms of
which the desired ratio $f_+/f_-$ can be expressed, as will be shown
in the next section.

The renormalization scheme is established in terms of the bare
$(n,l)$-vertex functions
$\Gamma_0^{(n,l)}(\{p;q\};m_0,g_0)$,
with $\{p;q\}=\{p_1,\ldots,p_n;q_1,\ldots,q_l\}$.
They emerge from the expectation values
$\left\langle\phi_0(x_1)\cdots\phi_0(x_n)\,\frac{1}{2}\phi_0^2(y_1)
\cdots\frac{1}{2}\phi_0^2(y_l)\right\rangle_c$ after
Legendre and Fourier transformation.
The renormalized mass (\ref{eq100}) is written as
\be
m^2_R=-Z_3\Gamma_0^{(2,0)}(0;m_0,g_0),\quad
\frac{1}{Z_3}:= -\frac{\partial\Gamma_0^{(2,0)}(p;m_0,g_0)}
                      {\partial p^2}\,\bigg|_{p^2=0}\,,
\label{eq120}
\ee
and in addition we define
\be
\frac{1}{Z_2}:= -\Gamma_0^{(2,1)}(\{0;0\};m_0,g_0)
              = -\frac{\partial}{\partial m_0^2}\,\Gamma_0^{(2,0)}
                          (0;m_0,g_0)\,.
\ee
The renormalization constants $Z_3$ and $Z_2$ play a prominent
r\^{o}le in the theory because they stand in close connection to
renormalization group functions and critical exponents.
For the renormalized coupling we use different settings
in each phase.
\subsubsection*{Symmetric phase}
As usual \cite{A84,MM94} we define a renormalized coupling
constant $g_R^{(4)}$ by the value of the four-point function for
vanishing external momenta:
\be
g_R^{(4)}:= -Z_3^2\Gamma_0^{(4,0)}(\{0\};m_0,g_0)\,.
\label{eq140}
\ee
Together with (\ref{eq120}) one can invert the relations between
$m_R$, $g_R^{(4)}$ on the one hand and $m_0$, $g_0$ on the other hand
so that the (in $D=3$) dimensionless renormalized coupling
\be
u_R:=\frac{g_R^{(4)}}{m_R}
\ee
becomes the natural expansion variable of our perturbation series.
Moreover with
\be
\frac{1}{Z_1}:= -\frac{1}{g_0}\,\Gamma_0^{(4,0)}(\{0\};m_0,g_0)
\ee
and (\ref{eq140}) we define for later purposes
\be
u:=\frac{g_0}{m_R}
  =u_R\,\frac{Z_1(u_R)}{Z_3^2(u_R)}\,,
\label{eq170}
\ee
where it is indicated that the dimensionless coupling $u$ is to be
read as a function of $u_R$.

To summarize, the renormalization is fixed by the conditions
\aeqn\bea
\Gamma_R^{(2,0)}(0;m_R,u_R)&=&-m_R^2 \\
\frac{\partial}{\partial p^2}\,\Gamma_R^{(2,0)}(p;m_R,u_R)\,
\bigg|_{p^2=0}&=&-1 \\
\Gamma_R^{(4,0)}(\{0\};m_R,u_R)&=&-m_R^{4-D}u_R\,=\,-g_R^{(4)} \\
\Gamma_R^{(2,1)}(\{0;0\};m_R,u_R)&=&-1 \\
\eea\reqn
for the renormalized $(n,l)$-vertex functions
\be
\Gamma_R^{(n,l)}(\{p;q\};m_R,u_R)
= [Z_3(u_R)]^{\frac{n}{2}-l}[Z_2(u_R)]^{l}
     \,\Gamma_0^{(n,l)}(\{p;q\};m_0,g_0)\,.
\ee
\subsubsection*{Broken symmetry phase}
In the phase of broken symmetry we follow \cite{LW87,M90} and
define the renormalized coupling constant $g_R$ by the vacuum
expectation value $v$ of the field $\phi$ in (\ref{eq30}).
If $G^{(1,0)}_c$ stands for the non-vanishing one-point function of the
field $\phi_0$ in (\ref{eq50}) one has
\be
v=v_0+G^{(1,0)}_c,\quad
v_0=\sqrt{3m_0^2/g_0}\,;\quad
v_R:=\frac{1}{\sqrt{Z_3}}\,v
\ee
with a renormalized vacuum expectation value $v_R$.
We set
\be
g_R:=\frac{3m_R^2}{v_R^2}\,,
\label{eq210}
\ee
and form a (in $D=3$) dimensionless renormalized coupling
according to
\be
u_R:=\frac{g_R}{m_R}\,.
\ee
Proceeding as in the previous subsection we find with
\be
Z_4:=\frac{m_0}{m_R}\left(1+\sqrt{\frac{g_0}{3m_0^2}}\,v_0
     \right)
\ee
and (\ref{eq210}) an expression for the dimensionless quantity
\be
u:=\frac{g_0}{m_R}
  =u_R\,\frac{Z_4^2(u_R)}{Z_3(u_R)}
\label{eq240}
\ee
in terms of renormalization constants, which again are assumed
to be (analytic) functions of $u_R$ alone.

The renormalization conditions fixing this scheme are:
\aeqn\bea
\Gamma_R^{(2,0)}(0;m_R,u_R)&=&-m_R^2 \\
\frac{\partial}{\partial p^2}\,\Gamma_R^{(2,0)}(p;m_R,u_R)\,
\bigg|_{p^2=0}&=&-1 \\
\frac{3 m_R^2}{v_R^2}&=&m_R^{4-D}u_R\, =\,g_R \\
\Gamma_R^{(2,1)}(\{0;0\};m_R,u_R)&=&-1\,.
\eea\reqn
%
\section{Determination of $f_+/f_-$}
Now we derive an expression for the amplitude ratio
$f_+/f_-$ in terms of analytic and dimensionless functions
calculable by renormalized perturbation theory.

First of all we have to realize the temperature dependence in
the parameters of this theory.
At first sight the occurrence of spontaneous symmetry breaking
in the Lagrangian (\ref{eq30}) suggests $t\propto m_0^2$, where
according to (\ref{eq20}) and (\ref{eq30})
\aeqn\bea
m_{0^+}^2&=&m_0^2 \\
m_{0^-}^2&=&-2m_0^2\,.
\eea\reqn
But when including the perturbative corrections one observes
a mass shift by an amount $m_{0c}^2$ in the temperature
variable, which is perturbatively not calculable \cite{BB85,BB87}.
(This corresponds to the fact that for $D<4$ mean field
theory is no longer valid.)
Distinguishing the different phases we have
\aeqn\bea
\frac{1}{m_{R^+}}=\xi_{+}\sim f_{+}t_{+}^{-\nu},
& &t_+ = \Big(m_0^2-m^2_{0c}\Big)\Big|_{T>T_c}>0
\label{eq270}\\
\frac{1}{m_{R^-}}=\xi_-\sim f_-t_-^{-\nu},
& &t_- = -\Big(m_0^2-m^2_{0c}\Big)\Big|_{T<T_c}>0
\label{eq275}
\eea\reqn
with $t_+=t$ for $t>0$ and $t_-=-t$ for $t<0$.

In order to circumvent this problem we eliminate the bare mass
$m_0$ by introducing the functions
\be
F_{\pm}(u_{R^{\pm}}):=\frac{\partial m^2_{R^{\pm}}}
                           {\partial m^2_{0^{\pm}}}\,\bigg|_{g_0},
\label{eq280}
\ee
which are partial derivatives with respect to $m_0^2$ at fixed
$g_0$ and depend in renormalized perturbation theory on $u_{R^{\pm}}$
only.
With the identity
\be
\frac{\partial}{\partial t}\,\bigg|_{g_0}
=\frac{\partial}{\partial m_0^2}\,\bigg|_{g_0}
\ee
one finds
\aeqn\bea
\frac{\partial m^2_{R^+}}{\partial t_+}\,\bigg|_{g_0}
&=&\frac{\partial m^2_{R^+}}{\partial m_0^2}\,\bigg|_{g_0}
=\frac{\partial m^2_{R^+}}{\partial m_{0^+}^2}\,\bigg|_{g_0}
=F_+(u_{R^+}) \\
\frac{\partial m^2_{R^-}}{\partial t_-}\,\bigg|_{g_0}
&=&-\frac{\partial m^2_{R^-}}{\partial m_0^2}\,\bigg|_{g_0}
=2\,\frac{\partial m^2_{R^-}}{\partial m_{0^-}^2}\,\bigg|_{g_0}
=2F_-(u_{R^-})\,.
\eea\reqn
Then the differentiation of (\ref{eq270}) and (\ref{eq275}) yields
\be
2\,\frac{F_-(u_{R^-})}{F_+(u_{R^+})}
=\frac{\partial m_{R^-}^2/\partial t_-}
 {\partial m_{R^+}^2/\partial t_+}\,\bigg|_{g_0}
\sim \left(\frac{f_+}{f_-}\right)^2\left(\frac{t_-}{t_+}\right)
 ^{2\nu-1}.
\ee
On the other hand the definition of the correlation length as the
inverse renormalized mass implies
\be
\left(\frac{f_+}{f_-}\right)^2
\left(\frac{t_-}{t_+}\right)^{2\nu}
\sim \left(\frac{m_{R^-}}{ m_{R^+}}\right)^2 \,,
\ee
and one concludes
\be
2\,\frac{F_-(u_{R^-})}{F_+(u_{R^+})}
\sim \left(\frac{m_{R^-}}{ m_{R^+}}\right)^2\frac{t_+}{t_-}\,.
\label{eq330}
\ee
So far $t_+$ and $t_-$, as well as $m_{R^+}$ and $m_{R^-}$, are
independent of each other.
In order to specify the approach to the critical point from both sides
we choose pairs of points $(t_+,t_-)$ in the phase diagram such that
\be
m_{R^+}=m_{R^-}\,.
\ee
This leads to a (yet unknown) dependence
between $u_{R^+}$ and $u_{R^-}$.
Consequently (\ref{eq270}), (\ref{eq275}) and (\ref{eq330})
read
\be
f_+t_+^{-\nu} \sim f_-t_-^{-\nu},\quad
\frac{t_+}{t_-} \sim 2\,\frac{F_-(u_{R^-})}{F_+(u_{R^+})}\,,
\ee
and we combine these equations into the formula
\be
\frac{f_+}{f_-}
\sim \left[\,2\,\frac{F_-(u_{R^-})}{F_+(u_{R^+})}\,
 \right]^{\nu}.
\label{eq360}
\ee
The crucial point is now that one can express (\ref{eq360})
as a function of a single dimensionless renormalized coupling constant
$\bar{u}_R$, which is related to $u_{R^{\pm}}$ in a definite way.

As mentioned earlier, the critical theory is characterized by the
non-trivial fixed point $u_R^*$ of $u_R$.
This is equal to the non-vanishing zero of the renormalization group
$\beta$-function
\be
\beta(u_R):= m_R\,\frac{\partial}{\partial m_R}\,
                 \bigg|_{g_0}\,u_R
           =-\left(\frac{\partial}{\partial u_R}\,
                 \bigg|_{m_R}\,\ln(u)\right)^{-1},\quad D=3\,,
\label{eq370}
\ee
according to $u=u(u_R)=g_0/m_R$.
Thus, if we are looking for a new coupling $\bar{u}_R$, which is to be
used as a renormalized coupling in both phases, we have to
ensure that this coupling leads to the same $\beta$-function in
both phases.
Consequently the functional relation between $m_{R^+}$ and $\bar{u}_R$
in the symmetric phase must be the same as the one between $m_{R^-}$ and
$\bar{u}_R$ in the phase of broken symmetry.
In view of (\ref{eq370}) one can comply with this condition by
using the relations (\ref{eq170}) and (\ref{eq240}), which
represent perturbative expansions of $u_{\pm}$ in terms of
$u_{R^{\pm}}$:
\aeqn\bea
u_- &=& h_-(u_{R^-}) \\
u_+ &=& h_+(u_{R^+})\,.
\eea\reqn
One possibility is to identify $\bar{u}_R$ with $u_{R^+}$ in the
symmetric phase and to define it in the phase with broken symmetry by
means of
\be
u_- = h_+(\bar{u}_R)\,.
\label{eq390}
\ee
The second possibility considered in this work is to identify
$\bar{u}_R$ with $u_{R^-}$ in the phase with broken symmetry and to
define it in the symmetric phase by means of
\be
u_+ = h_-(\bar{u}_R)\,.
\label{eq400}
\ee
In both cases the $\beta$-functions relevant for the high- and
low-temperature phases
\be
\bar{\beta}_{\pm}(\bar{u}_R)
=-\left[\,\frac{\partial}{\partial\bar{u}_R}\,
  \bigg|_{m_{R^{\pm}}}\,\ln\Big(u_{\pm}(\bar{u}_R)\Big)\,
  \right]^{-1}
\ee
are equal to each other,
\be
\bar{\beta}_{+}(\bar{u}_R)=\bar{\beta}_{-}(\bar{u}_R)\,.
\ee
In the first case their common fixed point value is
\be
\bar{u}_R^*=u_{R^+}^*\,,
\ee
for the second choice it is
\be
\bar{u}_R^*=u_{R^-}^*\,.
\ee
With the definition
\be
\Phi(\bar{u}_R) :=
\frac{F_-\Big(u_{R^-}(\bar{u}_R)
 \Big)}{F_+\Big(u_{R^+}(\bar{u}_R)\Big)}\,,
\ee
we can evaluate (\ref{eq360}) at the critical point,
\be
\frac{f_+}{f_-}
=\left[ 2\,\Phi(\bar{u}_R^*) \right]^{\nu},
\label{eq460}
\ee
to get numerical values for the amplitude ratio $f_+/f_-$ up to a given
order in perturbation theory.
%
\section{Results}
We have calculated the renormalization functions necessary for the
procedure outlined in the previous section in perturbation theory up
to the order of two loops.
In the symmetric phase 7 massive Feynman graphs and in the phase with
broken symmetry 24 additional graphs have been evaluated analytically in
$D=3$ dimensions with dimensional regularization.
We omit the details of our calculations and present only the main
results.
In the symmetric phase we obtain the bare expansions
\aeqn\bea
m_{R^+}^2&=&m_{0^+}^2\left\{1-\frac{1}{8\pi}\,u_{0^+}
            +\frac{1}{64\pi^2}\,u_{0^+}^2\left[\,\frac{79}{162}
            -\frac{1}{3}\,B_+^{(div)}\,\right]+O(u_{0^+}^3)\right\}
\label{eq470}\\
g_R^{(4)}&=&g_0\left\{1-\frac{3}{16\pi}\,u_{0^+}+\frac{5}{162\pi^2}\,
            u_{0^+}^2+O(u_{0^+}^3)\right\} \\
u_{R^+}&=&u_{0^+}\left\{1-\frac{1}{8\pi}\,u_{0^+}+\frac{1}{64\pi^2}\,
          u_{0^+}^2\left[\,\frac{293}{216}+\frac{1}{6}\,B_+^{(div)}\,
          \right]+O(u_{0^+}^3)\right\},
\eea\reqn
which are inverted to
\aeqn\bea
u_{0^+}&=&u_{R^+}\left\{1+\frac{1}{8\pi}\,u_{R^+}
          +\frac{1}{64\pi^2}\,u_{R^+}^2\left[\,\frac{139}{216}
          -\frac{1}{6}\,B^{(div)}_{R^+}\,\right]+O(u_{R^+}^3)\right\} \\
m_{0^+}^2&=&m_{R^+}^2\left\{1+\frac{1}{8\pi}\,u_{R^+}
            +\frac{1}{64\pi^2}\,u_{R^+}^2\left[\,\frac{245}{162}
          +\frac{1}{3}\,B^{(div)}_{R^+}\,\right]+O(u_{R^+}^3)\right\} \\
u_+&=&u_{R^+}\left\{1+\frac{3}{16\pi}\,u_{R^+}
      +\frac{575}{20736\pi^2}\,u_{R^+}^2+O(u_{R^+}^3)\right\}.
\label{eq480}
\eea\reqn
In the broken symmetry phase one has
\aeqn\bea
m_{R^-}^2&=&m_{0^-}^2\left\{1+\frac{3}{64\pi}\,u_{0^-}
          +\frac{1}{64\pi^2}\,u_{0^-}^2\left[\,\frac{19525}{5184}
          +\frac{2}{3}\,B_-^{(div)}\,\right]+O(u_{0^-}^3)\right\} \\
g_R&=&g_0\left\{1-\frac{7}{32\pi}\,u_{0^-}
      +\frac{37835}{331776\pi^2}\,u_{0^-}^2
      +O(u_{0^-}^3)\right\} \\
u_{R^-}&=&u_{0^-}\left\{1-\frac{31}{128\pi}\,u_{0^-}
          +\frac{1}{64\pi^2}\,u_{0^-}^2\left[\,\frac{80125}{13824}
          -\frac{1}{3}\,B_-^{(div)}\,\right]+O(u_{0^-}^3)\right\}
\eea\reqn
and
\aeqn\bea
u_{0^-}&=&u_{R^-}\left\{1+\frac{31}{128\pi}\,u_{R^-}
          +\frac{1}{64\pi^2}\,u_{R^-}^2\left[\,\frac{23663}{13824}
          +\frac{1}{3}\,B^{(div)}_{R^-}\,\right]+O(u_{R^-}^3)\right\} \\
m_{0^-}^2&=&m_{R^-}^2\left\{1-\frac{3}{64\pi}\,u_{R^-}
            -\frac{1}{64\pi^2}\,u_{R^-}^2\left[\,\frac{45125}{10368}
           +\frac{2}{3}\,B^{(div)}_{R^-}\,\right]+O(u_{R^-}^3)\right\}\\
u_-&=&u_{R^-}\left\{1+\frac{7}{32\pi}\,u_{R^-}
      -\frac{2191}{165888\pi^2}\,u_{R^-}^2+O(u_{R^-}^3)\right\}.
\label{eq500}
\eea\reqn

The divergent pieces
\aeqn\bea
B_{\pm}^{(div)}&:=&
\frac{1}{\epsilon} -\ln\left(\frac{m^2_{0^{\pm}}}{4\pi}\right)
-\gamma +\frac{const.}{4\pi} +O(\epsilon) \\
B_{R^{\pm}}^{(div)}&:=&
\frac{1}{\epsilon} -\ln\left(\frac{m^2_{R^{\pm}}}{4\pi}\right)
-\gamma +\frac{const.}{4\pi} +O(\epsilon)
\eea\reqn
vanish after differentiation with respect to $m_{0^{\pm}}^2$.
So we get from (\ref{eq280}) and the above equations
the finite functions
\bea
F_+(u_{R^+})&=&1-\frac{1}{16\pi}\,u_{R^+}
               -\frac{1}{384\pi^2}\,u_{R^+}^2+O(u_{R^+}^3) \\
F_-(u_{R^-})&=&1+\frac{3}{128\pi}\,u_{R^-}-\frac{233}{49152\pi^2}
               \,u_{R^-}^2+O(u_{R^-}^3)\,.
\eea

As discussed in the previous section the new coupling constant
$\bar{u}_R$ is to be introduced now.
We consider both choices explained above.
\begin{enumerate}
\item Expansion in $u_{R^+}$:\\
      We determine $u_{R^-}$ as a function of $\bar{u}_R$ with the help
      of equation (\ref{eq390}). From
      (\ref{eq480}) and (\ref{eq500}) the result is
      \be
      u_{R^-}=\bar{u}_R\left(1-\frac{1}{32\pi}\,\bar{u}_R
              +\frac{9059}{165888\pi^2}\,\bar{u}_R^2
              +O(\bar{u}_R^3)\right).
      \ee
      Since $\bar{u}_R^*=u_{R^+}^*$ we identify $\bar{u}_R$
      with $u_{R^+}$.
      Then
      \be
      \Phi_+(u_{R^+})
      =1+\frac{11}{128\pi}\,u_{R^+}+\frac{41}{16384\pi^2}\,u_{R^+}^2
       +O(u_{R^+}^3)\,.
      \label{eq550}
      \ee
\item Expansion in $u_{R^-}$:\\
      We adjust $\bar{u}_R$ to the low temperature coupling $u_{R^-}$.
      The equations (\ref{eq400}), (\ref{eq480}) and (\ref{eq500})
      show that in this case $u_{R^+}$ depends on $\bar{u}_R$ via
      \be
      u_{R^+}=\bar{u}_R\left(1+\frac{1}{32\pi}\,\bar{u}_R
              -\frac{8735}{165888\pi^2}\,\bar{u}_R^2
              +O(\bar{u}_R^3)\right),
      \ee
      and $\bar{u}_R^*=u_{R^-}^*$ allows the identification
      $\bar{u}_R=u_{R^-}$.
      Consequently
      \be
      \Phi_-(u_{R^-})
      =1+\frac{11}{128\pi}\,u_{R^-}
       +\frac{85}{16384\pi^2}\,u_{R^-}^2+O(u_{R^-}^3)\,.
       \label{eq570}
       \ee
\end{enumerate}

In order to get numerical results for the amplitude ratio $f_+/f_-$
according to (\ref{eq460}) we need the values of $u_{R^{\pm}}^*$ and
$\nu$.
The high temperature fixed point $u_{R^+}^*$ and the index $\nu$ have
been calculated in the framework of renormalized perturbation theory in
$D=3$ dimensions in \cite{LZ80,BB85}.
Therefore the whole analysis only requires the present methods.
To improve the quality of the results we prefer, however, to make use of
additional information available about $u_{R^{\pm}}^*$ and $\nu$.
For the exponent $\nu$ there are recent Monte Carlo result in
\cite{BGHP92}.
The fixed points $u_{R^{\pm}}^*$ have been estimated from the longest
presently available high- and low-temperature series in \cite{S93}, and
for $u_{R^-}^*$ we also use an estimate cited in \cite{M90}.
To estimate the sensitivity of $f_+/f_-$ to these parameters we list the
values resulting from different choices in tables 1 and 2.

The functions $\Phi_{\pm}(u_{R^{\pm}})$, given in equations
(\ref{eq550}) and (\ref{eq570}), have been evaluated at the fixed points
with Pad\'{e} approximants.
By this we do not mean that the perturbative expansion is analytic,
which it is presumably not, but just take the different results as a
measure for the error due to the shortness of the series.
The application of Pad\'e-Borel methods, which we also tried, does not
yield any improvement.

%
\begin{table}[htb]
\begin{center}
\begin{tabular}{|c|c||c|c|c|}
\hline
\multicolumn{2}{|c||}{\rule[-3mm]{0mm}{8mm} Input}
& \multicolumn{3}{c|}{\rule[-3mm]{0mm}{8mm} Amplitude ratio $f_+/f_-$}\\
\hline
Exponent $\nu$ & Fixed point $u_{R^+}^*$
& [2,0]-Pad\'{e} & [1,1]-Pad\'{e} & [0,2]-Pad\'{e}\\
\hline\hline
& $23.73(8)^{\mbox{\scriptsize \cite{LZ80}}}$
& 2.2348 & 2.2663 & 2.0716\\ \cline{2-5}
\raisebox{1.5ex}[-1.5ex]{$0.6300(15)^{\mbox{\scriptsize \cite{LZ80}}}$}
& $24.56(10)^{\mbox{\scriptsize \cite{S93}}}$
& 2.2605 & 2.2956 & 2.0775\\
\hline
& 23.73(8) & 2.2178 & 2.2487 & 2.0570\\ \cline{2-5}
\raisebox{1.5ex}[-1.5ex]{$0.624(2)^{\mbox{\scriptsize \cite{BGHP92}}}$}
& 24.56(10) & 2.2431 & 2.2775 & 2.0630\\
\hline
\end{tabular}
\end{center}
\begin{center}
\parbox{8cm}{{\footnotesize Table 1: }{\footnotesize
               Ratio $f_+/f_-$ for given
               values of $\nu$ and $u_{R^+}^*$.}}
\end{center}
\end{table}
%
%
\begin{table}[htb]
\begin{center}
\begin{tabular}{|c|c||c|c|c|}
\hline
\multicolumn{2}{|c||}{\rule[-3mm]{0mm}{8mm} Input}
& \multicolumn{3}{c|}{\rule[-3mm]{0mm}{8mm} Amplitude ratio $f_+/f_-$}\\
\hline
Exponent $\nu$ & Fixed point $u_{R^-}^*$
& [2,0]-Pad\'{e} & [1,1]-Pad\'{e} & [0,2]-Pad\'{e}\\
\hline\hline
& $14.73(14)^{\mbox{\scriptsize \cite{S93}}}$
& 2.0122 & 2.0497 & 2.0329\\ \cline{2-5}
\raisebox{1.5ex}[-1.5ex]{0.6300(15)}
& $15.1(1.3)^{\mbox{\scriptsize \cite{M90}}}$
& 2.0255 & 2.0660 & 2.0546\\
\hline
& 14.73(14) & 1.9988 & 2.0356 & 2.0255\\ \cline{2-5}
\raisebox{1.5ex}[-1.5ex]{0.624(2)} & 15.1(1.3) & 2.0119 & 2.0517
& 2.0406\\
\hline
\end{tabular}
\end{center}
\begin{center}
\parbox{8cm}{{\footnotesize Table 2: }{\footnotesize
                Ratio $f_+/f_-$ for given values
                of $\nu$ and $u_{R^-}^*$.}}
\end{center}
\end{table}
%
We summarize the contents of the tables by extracting mean values of
$f_+/f_-$ for the high and low temperature fixed points:
\be
\left(\frac{f_+}{f_-}\right)_{ht}=2.18(12)\,,\quad
\left(\frac{f_+}{f_-}\right)_{lt}=2.03(4)\,.
\ee
The number given in brackets is the maximal variation owing to the
inputs of $u_{R^{\pm}}^*$, $\nu$ and the different Pad\'{e}
approximants.

The fact that the variation between the different Pad\'{e}
approximations in table 2 is much smaller than in table 1 indicates that
the low temperature renormalized coupling constant $u_{R^-}$ is the
better expansion variable for our series.
An examination of (\ref{eq550}) and (\ref{eq570}) confirms this
impression since the first and second order perturbative corrections to
the leading terms in the coupling $u_{R^-}$ are smaller than those in
the high temperature coupling $u_{R^+}$ (41\% and 8\% against 66\% and
9\%).
Therefore we tend to have more faith in the estimate resulting from the
expansion in $u_{R^-}$.

For the sake of completeness we add a remark about the perturbative
determination of the critical exponent $\nu$.
It is related to the renormalization group functions $\eta_i$,
for $i=1,2$ defined by
\be
\eta_{i^{\pm}}(u_{R^{\pm}})
:=\beta_{\pm}(u_{R^{\pm}})\,\frac{\partial}{\partial u_{R^{\pm}}}
      \,\bigg|_{g_0}\,\ln\Big(Z_{i^{\pm}}(u_{R^{\pm}})\Big),
\ee
through the equations \cite{A84}
\be
\frac{1}{\nu_{\pm}(u_{R^{\pm}})}
:= 2-\eta_{3^{\pm}}(u_{R^{\pm}})
       +\eta_{2^{\pm}}(u_{R^{\pm}})\,,\quad
\nu=\nu_{\pm}(u^*_{R^{\pm}})\,.
\ee
The renormalization constants $Z_i$, $i=1,2$, are calculated
straightforwardly from (\ref{eq470})-(\ref{eq500}).
The expansions of $\nu_{\pm}$, which are of the form
\be
\nu_{\pm}=\frac{1}{2}+O(u_{R^{\pm}})\,,
\ee
can be inserted into (\ref{eq460}) and evaluated at the fixed points
$u^*_{R^{\pm}}$ in the same way as before.
The expansion has the form
$f_+/f_-=\sqrt{2}+O(u_{R^{\pm}})$.
This is the typical structure for renormalized perturbation theory
because the higher order corrections in $u_{R^{\pm}}$ represent the
in $D<4$ non-negligible deviations from the known mean field value
$(f_+/f_-)_{\mbox{\scriptsize\it mf}}=\sqrt{2}$.

With this method we find $(f_+/f_-)_{ht}=2.22(6)$ and
$(f_+/f_-)_{lt}=1.86(6)$.
We must emphasize, however, that these results are strongly influenced
by the behaviour of the short series for $\nu_{\pm}$ and therefore are
certainly less reliable than the results above, where more precise
information about $\nu$ has been employed.
%
\section{Conclusion}
Our results for the universal ratio of correlation length amplitudes
indicate a value of
\be
f_+/f_- =2.04(4)\,.
\ee
The theoretical estimates in the literature are 1.91 from the
$\epsilon$-expansion \cite{BLZ76,BLZ74}, and 1.96(1) \cite{LF89},
1.94(3) \cite{S93} from high- and low-temperature expansions.
They are definitely below our number.
On the other hand, experimental values from binary fluids are
2.05(22), 2.22(5) \cite{KKG83} and 1.9(2), 2.0(4) \cite{PHA91}.

In conclusion it appears quite reasonable to us that the correct value
of the universal amplitude ratio of correlation lengths can lie
above 2.
The calculation of further orders and a detailed investigation of the
resulting (non convergent, but likely asymptotic) perturbation series
may lead to more accurate estimates.

Finally let us mention that with an appropriate modification
our methods are also qualified to determine the amplitude ratios
$(f_+/f_-)_{ph}$ of the `true' correlation length defined via the
physical mass, and $C_+/C_-$ of the susceptibility.
We have performed the calculation of $C_+/C_-$ in the two-loop
approximation, too.
The result is consistent with the estimate in \cite{BB85}; their
series from Feynman graphs in the symmetric phase is, however, longer.
The calculation of the `true' correlation length in the one-loop
approximation reveals that the amplitude ratio is only negligibly
different from the one considered here.

As a consistency check we have also reproduced the $\epsilon$-expansions
of the quantities considered in this work.

%
\end{document}